\documentstyle[11pt,epsfig]{article} 
\if@twoside \oddsidemargin 21pt \evensidemargin 59pt \marginparwidth 85pt
\else \oddsidemargin 0pt \evensidemargin 0pt
\fi
\renewcommand{\today}{II. Institute of Theoretical Physics,\\ University of
Stuttgart,\\ 70550 Stuttgart, Germany}                           
\begin{document}                                     
\title{Self-Organization Phenomena in Pedestrian Crowds}
\author{Dirk Helbing and P\'{e}ter Moln\'{a}r}
\renewcommand{\today}{II. Institute of Theoretical Physics,\\ University of
Stuttgart,\\[2mm] 70550 Stuttgart, Germany}                           
\maketitle  
\begin{abstract}                                           
Pedestrian crowds can very realistically be simulated with a social force
model which describes the different influences affecting individual pedestrian
motion by a few simple force terms. The model is able to reproduce the
emergence of several empirically observed collective patterns of motion.
These self-organization phenomena can be utilized for new flow optimization
methods which are indispensable for skilful town- and traffic-planning.
\end{abstract}

\section*{Introduction}

Although group dynamics has always been a fascinating field of research,
many phenomena of collective behavior patterns are still only qualitatively
understood. Apart from the complexity of social phenomena, one serious
problem is that many variables (quantities) which influence 
human behavior are hardly measurable. For this reason pedestrian crowds
are an ideal object of social research: All essential quantities like
places, speeds, and walking directions of pedestrians as well as locations
of obstacles and attractions etc. can be easily and exactly determined.
Moreover, a large amount of data has already been collected 
\cite{b3,weidm}.
\par
Pedestrian crowds have been studied for about three decades now
\cite{Oed,Old,Nav,b3,weidm}. 
The main goal of these studies was to develop guidelines
for planning and designing pedestrian facilities. These have usually
the form of regression relations \cite{b3} which are, however, not very
well suited for the prediction of pedestrian flows in pedestrian precincts
or buildings with an exceptional architecture. Therefore, a number of
simulation models have been proposed, e.g. {\em queueing models} 
\cite{b4,Ash,que1,b5} and models for the route choice behavior of pedestrians
\cite{b6}. 
\par
None of these approaches adequately takes into account the
{\em self-organization effects} occuring in pedestrian crowds. 
These may, however, lead to unexpected obstructions due to mutual disturbances
of pedestrian flows. The different kinds of spatio-temporal collective
patterns of motion formed by pedestrian crowds even have not been investigated
in detail yet. One reason for this may be the sensitive dependence of the
emerging pedestrian flows on the geometrical shape of pedestrian facilities.
\par
In the following section we will describe some
of our observations concerning pedestrian motion. Afterwards a new
model for the behavior of pedestrians and their interactions with the
developed environment and other pedestrians will be presented. 
By means of computer simulations we will 
demonstrate that this model is able to explain the formation
of the observed self-organization phenomena. 
Finally, it is shown that self-organization
effects can be utilized for an optimization of pedestrian flows. 
That means, computer simulations like the ones presented can 
provide a powerful tool
for designing and planning pedestrians facilities like subway or railway
stations, pedestrian precincts, shopping malls, or big buildings.

\section*{Observations}

We have studied pedestrian motion for several years and evaluated a number of
quick-motion video films \cite{Arns}. Despite the sometimes more or less
`chaotic' appearance of individual pedestrian behavior, some regularities
can be found. The following results mainly apply to purposeful pedestrians
who steer for a certain destination. They are not valid for aimless pedestrians
like tourists who stroll around (showing other rules of motion, cf.
\cite{b9}). In addition, children move more irregular than described below
since they must learn optimal strategies of motion first which are used
more or less automatically later.
\par
Our observations can be summarized as follows:
\begin{itemize}
\item[1.] Pedestrians normally choose the shortest route to their
next destination which has therefore the shape of a polygon. If alternative
routes have the same length, a pedestrian prefers the one where he/she can
go straight ahead as long as possible and change the direction as late as
possible provided that the alternative route is not more attractive (due to 
less noise, more light, friendlier environment, less waiting time at traffic
lights, etc.).
A pedestrian feels a strong aversion to taking detours or moving
opposite to the desired walking direction even if the direct way is crowded.
\item[2.] Pedestrians prefer to walk with an individual desired speed which
corresponds to the comfortable walking speed as long as it is not necessary to
move faster in order to reach the 
destination in time. The desired speeds within
pedestrian crowds are {\sc Gaussian} distributed \cite{Nav,Hen1,Hen2,Hen3}.
\item[3.] Pedestrians keep a certain distance to other pedestrians and borders
(of streets, walls, and obstacles). This distance is the smaller the more a
pedestrian hurries, and it decreases with growing pedestrian density. 
\par
Resting individuals (waiting on a railway platform for a train, sitting in 
a dining hall, or lying at a beach) are uniformly distributed
over the available area if there are no 
aquaintances among the individuals. Pedestrian density increases (i.e. 
interpersonal distances lessen) around particularly attractive places.
It decreases with growing velocity variance (e.g. on a dance floor, cf.
\cite{b8,sfb}). Individuals knowing each other may form groups which are
entities that behave similar to single pedestrians. Group
sizes are {\sc Poisson} distributed \cite{Col1,Col2,b9}.
\item[4.] Pedestrians normally do not reflect their behavioral strategy in 
every situation anew but act more or less automatically (as an experienced car 
driver does). This becomes obvious when pedestrians cause delays or
obstructions, e.g. by already entering an elevator or underground even
though others still try to get off.
\end{itemize}
Additionally, we found that, at medium and high pedestrian densities,
the motion of pedestrian crowds shows some striking analogies
with the motion of gases and fluids. This includes 
\begin{itemize}
\item[1.] the similarity of footprints in snow with streamlines of fluids, 
\item[2.] the existence of a condensed state of pedestrian crowds in which 
pedestrians cannot move freely any more,
\item[3.] `viscous fingering' \cite{fin1,fin2} at the borderline between 
opposite streams of motion,
\item[4.] the development of river-like streams through {\em waiting} 
pedestrian crowds, 
\item[5.] and the formation of pedestrian-free bubbles.
\end{itemize}
The paper will now focus on the explanation of some 
frequently observable self-organization phenomena which have not been
modelled by other theories yet:
\begin{itemize}
\item[1.] In crowds of pedestrians moving into opposite directions, lanes of
uniform walking direction are formed if the pedestrian density exceeds a
critical value. 
\item[2.] Oscillatory changes of the walking direction develop
at narrow passages (corridors, staircases, or doors).
The average oscillation frequency
increases with growing width and decreasing length of the passage.
\end{itemize}

\section*{The Social Force Model for Pedestrian Motion}

The idea to describe pedestrian crowds by a {\em fluid-dynamic} model goes 
back to {\sc Henderson} \cite{b7}. However, since his model bases on 
unrealistic assumptions like energy and momentum conservation during
pedestrian interactions, an improved fluid-dynamic model was
proposed \cite{b8} which is derivable from a pedestrian-specific gaskinetic
model \cite{b8,diss}.
\par
However, 
for practical applications 
a direct simulation of {\em individual} pedestrian behavior is favourable
since a numerical solution of the fluid-dynamic equations is very difficult.
As a consequence, current research concentrates on 
the {\em microsimulation} of 
pedestrian crowds. In this connection, a {\em social force model} of individual
pedestrian dynamics was recently suggested \cite{b10,SFB}. 
A simple forerunner of this kind of model 
was proposed by {\sc Gipps} and {\sc Marksj\"o} \cite{b12}. 
\par
The social force concept of pedestrian motion can shortly be described
as follows:
Pedestrians are {\em used} to the situations they are normally confronted
with. Their behavioral strategies
are determined by their experience which reaction to 
a certain stimulus (situation) will be the {\em best}.
Therefore, their reactions are usually rather {\em `automatic'} and well
predictable. 
\par
We will see that pedestrian motion can be described as if it would
be governed by a `social force'. 
The {\em social force} $\vec{f}_\alpha$ represents
the different influences (of environment and other pedestrians) on the behavior
of a pedestrian $\alpha$. It determines the temporal change 
$d\vec{v}_\alpha/dt$ of his/her 
{\em actual velocity} $\vec{v}_\alpha = d\vec{r}_\alpha/dt$ 
together with a {\em fluctuation term} 
which delineates random behavioral variations (arising from accidental
or deliberate deviations from the usual rules of motion):
\begin{equation}
 \frac{d\vec{v}_\alpha}{dt} = \vec{f}_\alpha(t) + \mbox{\em fluctuations.}
\end{equation}
However, in contrast to physical forces, the social force is not excerted
on a pedestrian by his/her environment. Instead it reflects the concrete
{\em psychological motivation} 
to act causing pedestrian $\alpha$ to produce
an acceleration or deceleration force of strength $\vec{f}_\alpha(t)$ 
himself/herself. (For a foundation of the social force concept cf. Refs.
\cite{Lew,PhysA,Kluw}.)
\par
The social force $\vec{f}_\alpha(t)$ consists of several force terms which 
correspond to the different influences simultaneously affecting the
behavior of pedestrian $\alpha$:
\begin{equation}
  \vec{f}_\alpha(t) =  \vec{f}_\alpha^{\,0}(\vec{v}_\alpha,
 v_\alpha^0\vec{e}_\alpha)
 + \vec{f}_{\alpha B}(\vec{r}_\alpha - \vec{r}_B^{\,\alpha})
 + \sum_{\beta(\ne \alpha)} \vec{f}_{\alpha \beta} (
 \vec{r}_\alpha - \vec{r}_\beta) 
 + \sum_{i} \vec{f}_{\alpha i} (
 \vec{r}_\alpha - \vec{r}_i,t) \, .
\end{equation}
\begin{itemize}
\item[1.] Each pedestrian wants to walk with an individual {\em desired
speed} $v_\alpha^0$ in the direction $\vec{e}_\alpha$ of his/her next 
destination. Deviations of the {\em actual velocity} $\vec{v}_\alpha$ from the
{\em desired velocity} $\vec{v}_\alpha^{\,0} = v_\alpha^0 \vec{e}_\alpha$ due 
to disturbances (by obstacles or avoidance maneuvers) are corrected within
the so-called {\em `relaxation time'} $\tau_\alpha$:
\begin{equation}
  \vec{f}_\alpha^{\,0}(\vec{v}_\alpha,v_\alpha^0\vec{e}_\alpha)
= \frac{1}{\tau_\alpha} (v_\alpha^0 \vec{e}_\alpha - \vec{v}_\alpha ) \, .
\end{equation}
In order to compensate for delays, the desired speed $v_\alpha^0(t)$
is often increased in the course of time.
\item[2.] Pedestrians keep some distance from 
borders in order to avoid the risk of getting hurt. 
The closer the border is the more incomfortable a pedestrian feels. This effect
can be described by a repulsive force $\vec{f}_{\alpha B}$
which decreases monotonously with
the distance $\|\vec{r}_{\alpha} - \vec{r}_B^{\,\alpha}\|$ between the
place $\vec{r}_\alpha(t)$ of pedestrian $\alpha$ and
the nearest piece of the border. In the the simplest case this force can be
expressed in terms of a repulsive potential $V_{B}(r)$:
\begin{equation}
 \vec{f}_{\alpha B}(\vec{r}_\alpha - \vec{r}_B^{\,\alpha})
 := - \nabla_{\vec{r}_{\alpha}}
 V_{B} (\|\vec{r}_{\alpha} - \vec{r}_B^{\,\alpha}\|) \, .
\end{equation}
\par
Similar repulsive force terms $\vec{f}_{\alpha\beta}(\vec{r}_{\alpha} - 
\vec{r}_{\beta})$ can describe that each pedestrian $\alpha$ keeps a
situation-dependent distance to the 
other pedestrians $\beta$. This reflects the 
tendency to respect a {\em private sphere (`territorial effect')} and helps to
avoid collisions in cases of sudden velocity changes. 
\item[3.] Pedestrians show a certain joining behavior. For example,
families, friends, or tourist parties often move in groups. 
In addition, pedestrians are sometimes attracted 
by window displays, sights, special performances (street artists), or 
unusual events at places $\vec{r}_i$. Both situations can be modelled by
time-dependent attractive forces $\vec{f}_{\alpha i}(\vec{r}_{\alpha} - 
\vec{r}_{i},t)$ in a similar way like the repulsive effects but with an
opposite sign. 
\end{itemize}
For a more detailed discussion and the concrete mathematical specification of
the force terms cf. Ref. \cite{b10}.
                                                               
\section*{Simulation of Pedestrian Crowds and Self-Organization}

The social force model of pedestrian
dynamics has been simulated on a computer for a large number of interacting
pedestrians confronted with different situations. Despite the fact that
the proposed model is very simple it describes a lot of observed
phenomena very realistically. Especially, under certain conditions
the {\em self-organization} of collective behavioral patterns can be
observed. `Self-organization' means that these patterns are not externally
planned, prescribed, or organized, e.g. by traffic signs, laws, or behavioral
conventions. Instead, the spatio-temporal patterns emerge due to the non-linear
interactions of pedestrians. Our model (according to which 
individuals behave rather automatically) can explain the self-organized 
patterns described in the following without assuming strategical
considerations, communication, or imitative behavior of pedestrians:
\begin{itemize}
\item[1.] Above a critical pedestrian density our simulations reproduce the
empirically observed formation of lanes consisting of pedestrians with the
same desired walking direction (cf. Fig. 1). 
These lanes are dynamically varying. Their number
depends on the width of the street and on pedestrian density.
\par
The conventional interpretation of lane formation assumes that pedestrians
tend to walk on the side which is prescribed in vehicular traffic. However,
our model can explain lane formation although it does not assume a preference
for {\em any} side (it is completely symmetrical with respect to the left-hand
and the right-hand side). The mechanism of lane formation is that 
motion is more efficient for a pedestrian who follows other pedestrians with
the same walking direction since avoidance maneuvers are less frequently
necessary, then. 
\item[2.] In our simulations oscillatory changes of the walking direction
at narrow passages are also observed (cf. Fig. 2). The conventional 
interpretation for a change of the walking direction is that, after some time,
a pedestrian does the kindness of giving 
precedence to a waiting pedestrian with
an opposite walking direction. This, however, cannot explain the increase of 
oscillation frequency with passage width.
\par
The mechanism leading to alternating flows is the following: Once a
pedestrian is able to pass the narrowing (door, staircase, etc.), pedestrians
with the same walking direction can easily follow which is particularly clear
for long passages. By this the number and `pressure'
of waiting and pushing pedestrians becomes less than
on the other side of the narrowing where, consequently, 
the chance to occupy the passage
grows. This leads to a deadlock situation after some time which is followed
by a change of the walking direction. Capturing the passage is easier if it
is broad and short so that the walking direction changes more frequently, then.
\item[3.] At intersections our simulations show the temporary emergence of
unstable roundabout traffic with an alternating rotation direction
(cf. Fig. 3). (After the discovery of this effect in our computer 
simulations it was also empirically confirmed.) 
\par
Roundabout traffic is connected with small detours but decreases the
frequency of necessary deceleration, stopping, and avoidance maneuvers
considerably so that pedestrian motion becomes more efficient on average.
\end{itemize}

\section*{Optimization of Pedestrian Facilities}

The emerging pedestrian flows decisively depend on pedestrian density
and the geometry of pedestrian facilities. 
They can be simulated on a computer already in the planning phase of
pedestrian facilities. The configuration
and shape of the facilities can be systematically varied 
(e.g. by {\em evolutionary algorithms} \cite{Rech}) and evaluated on the
basis of particular mathematical {\em performance measures}. For example, the
{\em efficiency measure} 
\begin{equation}
 E := \frac{1}{N} \sum_{\alpha} 
 \overline{\frac{\vec{v}_\alpha \cdot \vec{e}_\alpha}
  {v_\alpha^0}} \qquad (0 \le E \le 1)
\end{equation}
(where $N$ is the number of pedestrians $\alpha$ and the bar denotes
a time average) calculates the mean value of the velocity
component into the desired direction of motion in relation to the desired
walking speed. The {\em uncomfortableness measure} 
\begin{equation}
 U := \frac{1}{N} \sum_{\alpha} \frac{\overline{(\vec{v}_\alpha
 - \overline{\vec{v}_\alpha})^2}}{\overline{(\vec{v}_\alpha)^2}}
 = \frac{1}{N} \sum_{\alpha} \left( 1 - \frac{\overline{\vec{v}_\alpha}^2}
 {\overline{(\vec{v}_\alpha)^2}} \right) \qquad (0 \le U \le 1)
\end{equation}
reflects the frequency and degree of sudden velocity changes, i.e. 
the level of discontinuity of
walking due to necessary avoidance maneuvers. Hence, the optimal
configuration regarding the pedestrian requirements is the one
with the highest values of efficiency and comfortableness.
\par
During the optimization procedure, some or all of the following 
can be varied:
\begin{itemize}
\item[1.] the location and form of planned buildings,
\item[2.] the arrangement of walkways, entrances, exits, stairs, 
elevators, escalators, and corridors,
\item[3.] the shape of rooms, corridors, entrances, and exits,
\item[4.] the function and time schedule of room usage. 
\end{itemize}
The proposed optimization procedure can not only be applied 
to the design of new
pedestrian facilities but also to a reduction of existing bottlenecks
by suitable modifications. Here, we give two examples:
\begin{itemize}
\item[1.] A broader door does not necessarily increase the efficiency of
pedestrian flow through it. It may rather lead to more frequent changes
of the walking direction which are connected with an increase of deadlook 
situations. Therefore, two doors are much more efficient than one single
door with double width. By self-organization \cite{Kluw}
each door is used by one walking direction, then (cf. Fig. 4).
\item[2.] Oscillatory changes of the walking direction and periods of
standstill in between also occur when different flows {\em cross} each other.
The loss of efficiency caused by this
can be reduced by psychological guiding methods or railings initializing
roundabout traffic. Roundabout traffic can already be induced and stabilized
by planting a tree in the middle of a crossing. 
In our simulations this increased efficiency up to 13\%.
\end{itemize} 
The complex interaction between various flows can lead to completely
unexpected results due to the {\em non-linearity 
of dynamics}. (A very impressive
and surprising result of evolutionary form optimization is presented in
Ref. \cite{b11}.) That means, planning of pedestrian facilities with
conventional methods cannot guarantee the avoidance of
big jams, serious obstructions, and catastrophic 
blockages (especially in emergency situations).
In contrast, a skilful flow optimization not only enhances efficiency
but also saves space that can be used for kiosks, benches, or other purposes. 
                           
\section*{Summary and Outlook}

It has been shown that pedestrian motion
can be described by a simple social force model for individual
pedestrian behavior. Computer simulations of pedestrian crowds demonstrated 
1.\ the development of lanes consisting of pedestrians who walk into the
same direction, 2.\ oscillatory changes of the walking direction at
narrow passages, 3.\ the spontaneous formation of roundabout traffic at
crossings. These spatio-temporal patterns are self-organized.
They arise due to the {\em nonlinear
interactions} of pedestrians. 
\par
Presently, the social force model is extended by a
model for the route choice behavior of pedestrians which facilitates the 
simulation of complex systems of ways. The sequence
of destinations a pedestrian visits can be simulated by a 
{\em discrete choice model}
(cf. \cite{b6,diss}). Another extension will allow the simulation of 
{\em trail formation} \cite{SFB}.
\par
Since the variables of pedestrian motion are easily measurable,
pedestrian models are comparable with empirical data.
Moreover, the study of pedestrian 
crowds leads to a deeper insight into the mechanisms underlying 
the self-organization phenomena in social systems. Therefore, pedestrian models
are an ideal starting point for the development of 
quantitative behavioral models concerning group dynamics \cite{Kluw}, opinion
formation \cite{MathS}, and other social processes.

\clearpage
\begin{figure}[htbp]
\begin{center}
    \epsfig{height=10cm, bbllx=125pt, bblly=128pt, bburx=365pt,
      bbury=719pt, angle=-90, file=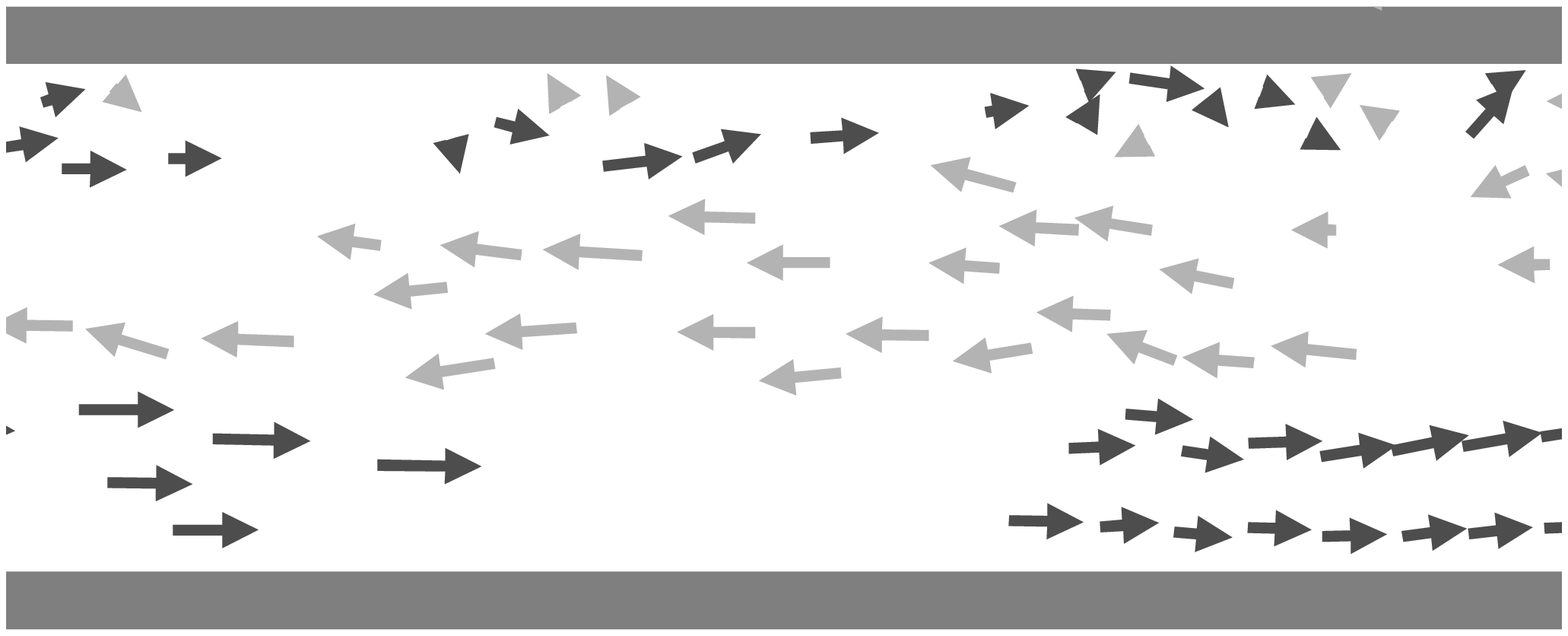}
\end{center}
\caption[]{Three lanes have emerged by the interactions between pedestrians
with opposite walking directions. (Each arrow represents the actual velocity
of one pedestrian.)}
\end{figure}
\begin{figure}[htbp]
\begin{center}
    \epsfig{height=11cm, bbllx=133pt, bblly=128pt, bburx=347pt,
      bbury=719pt, angle=-90, file=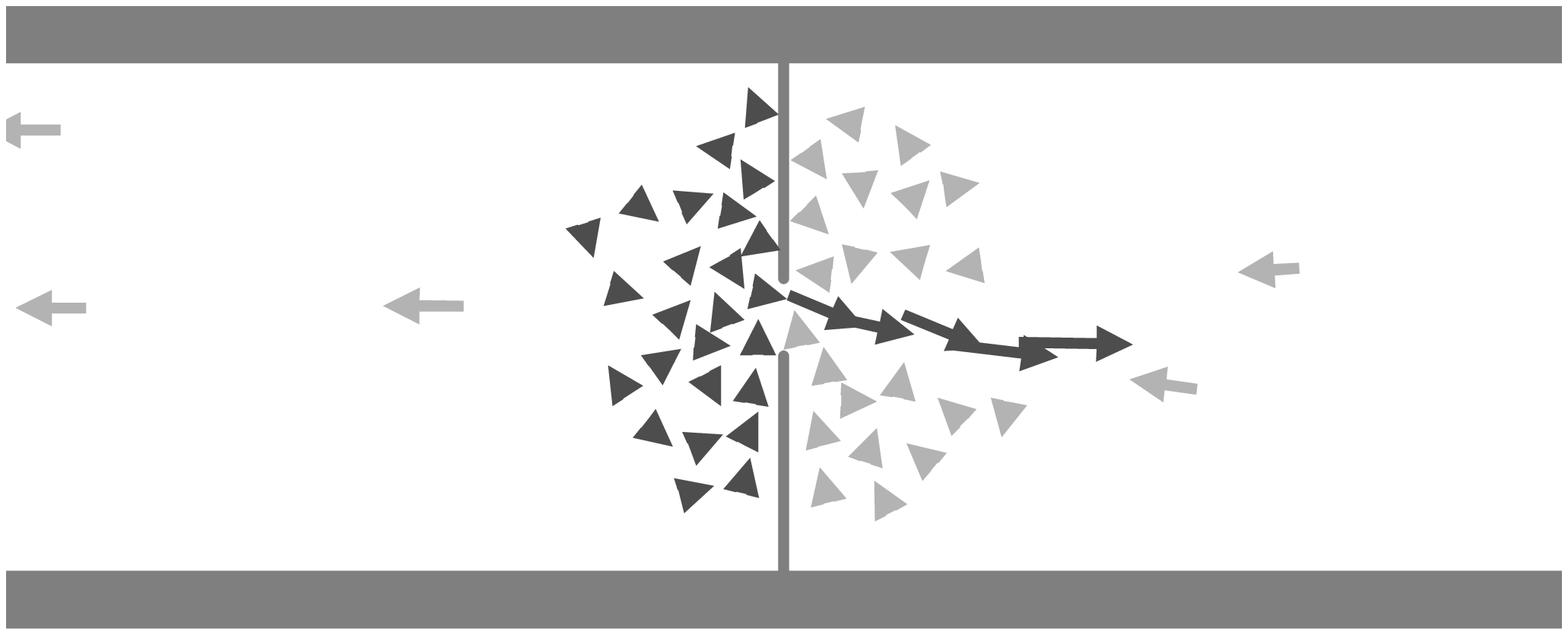} \\[1cm]
    \epsfig{height=11cm, bbllx=133pt, bblly=128pt, bburx=347pt,
      bbury=719pt, angle=-90, file=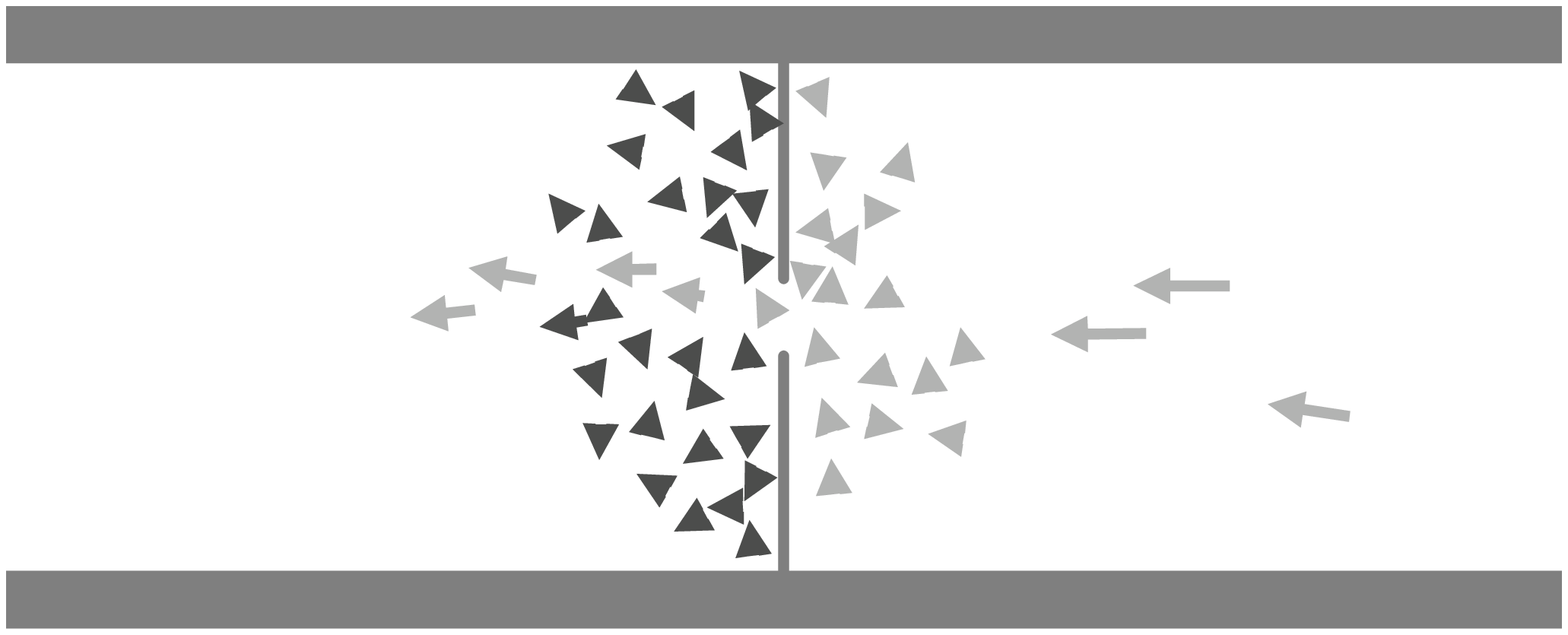} \\[7mm]
\end{center}
\caption[]{The above snapshots of two succeeding situations shows that the
walking direction at a narrow passage changes in an oscillatory way.}
\end{figure} 
\begin{figure}[htbp]
\begin{center}
    \epsfig{height=11cm, angle = -90, 
      bbllx=19pt, bblly=151pt, bburx=490pt, bbury=696pt,
      file=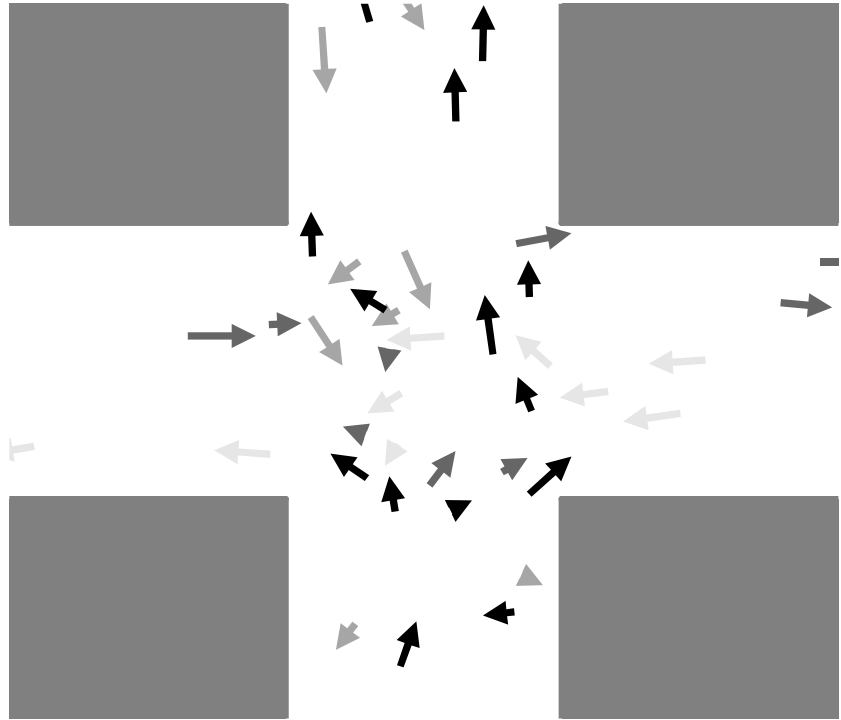}
\end{center}
\caption[]{At intersections the self-organization of unstable roundabout
traffic can be observed.}
\end{figure} 
\begin{figure}[htbp]
\begin{center}
    \leavevmode
    \epsfig{height=11.7cm, bbllx=281pt, bblly=115pt, bburx=471pt,
      bbury=734pt, angle=-90, file=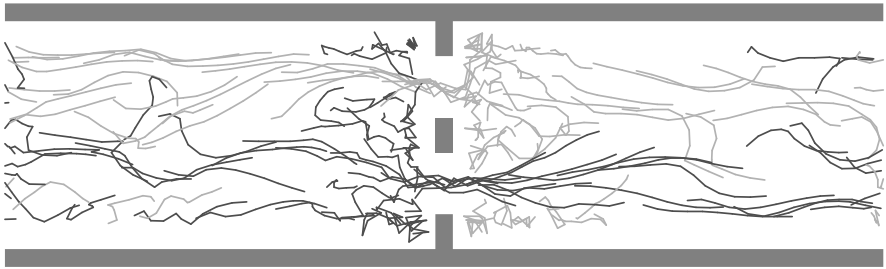}
\end{center}
\caption[]{The pedestrian trajectories show that, in the case of two 
alternative passages, each door is occupied by one walking direction
for a long time period.} 
\end{figure} 
\end{document}